\RequirePackage{lineno}
\documentclass[aps,showpacs,twocolumn,groupedaddress,superscriptaddress]{revtex4}
\usepackage{graphicx}  % needed for figures
\usepackage{dcolumn}   % needed for some tables
\usepackage{bm}        % for math
\usepackage{amssymb}   % for math
\usepackage{slashed}
\usepackage{amsmath}
\newcommand{\MET}{\slashed{E}_T}
\newcommand{\ppbar}{p\overline{p}}
\begin{document}
%\MakeLineNo
% The following information is for internal review,
% please remove them for submission
%\leftline{To be submitted to PRL}

% the following line is for submission, including submission to the arXiv!!
\hspace{5.2in} \mbox{FERMILAB-PUB-10-262-E}

\title{Search for sneutrino production in $\boldsymbol{e\mu}$ final states in 5.3~fb$^{-1}$ of $\boldsymbol{p\overline{p}}$ collisions at $\boldsymbol{\sqrt{s} =1.96}$~TeV }
%\author{D0 collaboration}
%
\affiliation{Universidad de Buenos Aires, Buenos Aires, Argentina}
\affiliation{LAFEX, Centro Brasileiro de Pesquisas F{\'\i}sicas, Rio de Janeiro, Brazil}
\affiliation{Universidade do Estado do Rio de Janeiro, Rio de Janeiro, Brazil}
\affiliation{Universidade Federal do ABC, Santo Andr\'e, Brazil}
\affiliation{Instituto de F\'{\i}sica Te\'orica, Universidade Estadual Paulista, S\~ao Paulo, Brazil}
\affiliation{Simon Fraser University, Vancouver, British Columbia, and York University, Toronto, Ontario, Canada}
\affiliation{University of Science and Technology of China, Hefei, People's Republic of China}
\affiliation{Universidad de los Andes, Bogot\'{a}, Colombia}
\affiliation{Charles University, Faculty of Mathematics and Physics, Center for Particle Physics, Prague, Czech Republic}
\affiliation{Czech Technical University in Prague, Prague, Czech Republic}
\affiliation{Center for Particle Physics, Institute of Physics, Academy of Sciences of the Czech Republic, Prague, Czech Republic}
\affiliation{Universidad San Francisco de Quito, Quito, Ecuador}
\affiliation{LPC, Universit\'e Blaise Pascal, CNRS/IN2P3, Clermont, France}
\affiliation{LPSC, Universit\'e Joseph Fourier Grenoble 1, CNRS/IN2P3, Institut National Polytechnique de Grenoble, Grenoble, France}
\affiliation{CPPM, Aix-Marseille Universit\'e, CNRS/IN2P3, Marseille, France}
\affiliation{LAL, Universit\'e Paris-Sud, CNRS/IN2P3, Orsay, France}
\affiliation{LPNHE, Universit\'es Paris VI and VII, CNRS/IN2P3, Paris, France}
\affiliation{CEA, Irfu, SPP, Saclay, France}
\affiliation{IPHC, Universit\'e de Strasbourg, CNRS/IN2P3, Strasbourg, France}
\affiliation{IPNL, Universit\'e Lyon 1, CNRS/IN2P3, Villeurbanne, France and Universit\'e de Lyon, Lyon, France}
\affiliation{III. Physikalisches Institut A, RWTH Aachen University, Aachen, Germany}
\affiliation{Physikalisches Institut, Universit{\"a}t Freiburg, Freiburg, Germany}
\affiliation{II. Physikalisches Institut, Georg-August-Universit{\"a}t G\"ottingen, G\"ottingen, Germany}
\affiliation{Institut f{\"u}r Physik, Universit{\"a}t Mainz, Mainz, Germany}
\affiliation{Ludwig-Maximilians-Universit{\"a}t M{\"u}nchen, M{\"u}nchen, Germany}
\affiliation{Fachbereich Physik, Bergische  Universit{\"a}t Wuppertal, Wuppertal, Germany}
\affiliation{Panjab University, Chandigarh, India}
\affiliation{Delhi University, Delhi, India}
\affiliation{Tata Institute of Fundamental Research, Mumbai, India}
\affiliation{University College Dublin, Dublin, Ireland}
\affiliation{Korea Detector Laboratory, Korea University, Seoul, Korea}
\affiliation{CINVESTAV, Mexico City, Mexico}
\affiliation{FOM-Institute NIKHEF and University of Amsterdam/NIKHEF, Amsterdam, The Netherlands}
\affiliation{Radboud University Nijmegen/NIKHEF, Nijmegen, The Netherlands}
\affiliation{Joint Institute for Nuclear Research, Dubna, Russia}
\affiliation{Institute for Theoretical and Experimental Physics, Moscow, Russia}
\affiliation{Moscow State University, Moscow, Russia}
\affiliation{Institute for High Energy Physics, Protvino, Russia}
\affiliation{Petersburg Nuclear Physics Institute, St. Petersburg, Russia}
\affiliation{Stockholm University, Stockholm and Uppsala University, Uppsala, Sweden }
\affiliation{Lancaster University, Lancaster LA1 4YB, United Kingdom}
\affiliation{Imperial College London, London SW7 2AZ, United Kingdom}
\affiliation{The University of Manchester, Manchester M13 9PL, United Kingdom}
\affiliation{University of Arizona, Tucson, Arizona 85721, USA}
\affiliation{University of California Riverside, Riverside, California 92521, USA}
\affiliation{Florida State University, Tallahassee, Florida 32306, USA}
\affiliation{Fermi National Accelerator Laboratory, Batavia, Illinois 60510, USA}
\affiliation{University of Illinois at Chicago, Chicago, Illinois 60607, USA}
\affiliation{Northern Illinois University, DeKalb, Illinois 60115, USA}
\affiliation{Northwestern University, Evanston, Illinois 60208, USA}
\affiliation{Indiana University, Bloomington, Indiana 47405, USA}
\affiliation{Purdue University Calumet, Hammond, Indiana 46323, USA}
\affiliation{University of Notre Dame, Notre Dame, Indiana 46556, USA}
\affiliation{Iowa State University, Ames, Iowa 50011, USA}
\affiliation{University of Kansas, Lawrence, Kansas 66045, USA}
\affiliation{Kansas State University, Manhattan, Kansas 66506, USA}
\affiliation{Louisiana Tech University, Ruston, Louisiana 71272, USA}
\affiliation{University of Maryland, College Park, Maryland 20742, USA}
\affiliation{Boston University, Boston, Massachusetts 02215, USA}
\affiliation{Northeastern University, Boston, Massachusetts 02115, USA}
\affiliation{University of Michigan, Ann Arbor, Michigan 48109, USA}
\affiliation{Michigan State University, East Lansing, Michigan 48824, USA}
\affiliation{University of Mississippi, University, Mississippi 38677, USA}
\affiliation{University of Nebraska, Lincoln, Nebraska 68588, USA}
\affiliation{Rutgers University, Piscataway, New Jersey 08855, USA}
\affiliation{Princeton University, Princeton, New Jersey 08544, USA}
\affiliation{State University of New York, Buffalo, New York 14260, USA}
\affiliation{Columbia University, New York, New York 10027, USA}
\affiliation{University of Rochester, Rochester, New York 14627, USA}
\affiliation{State University of New York, Stony Brook, New York 11794, USA}
\affiliation{Brookhaven National Laboratory, Upton, New York 11973, USA}
\affiliation{Langston University, Langston, Oklahoma 73050, USA}
\affiliation{University of Oklahoma, Norman, Oklahoma 73019, USA}
\affiliation{Oklahoma State University, Stillwater, Oklahoma 74078, USA}
\affiliation{Brown University, Providence, Rhode Island 02912, USA}
\affiliation{University of Texas, Arlington, Texas 76019, USA}
\affiliation{Southern Methodist University, Dallas, Texas 75275, USA}
\affiliation{Rice University, Houston, Texas 77005, USA}
\affiliation{University of Virginia, Charlottesville, Virginia 22901, USA}
\affiliation{University of Washington, Seattle, Washington 98195, USA}
\author{V.M.~Abazov} \affiliation{Joint Institute for Nuclear Research, Dubna, Russia}
\author{B.~Abbott} \affiliation{University of Oklahoma, Norman, Oklahoma 73019, USA}
\author{M.~Abolins} \affiliation{Michigan State University, East Lansing, Michigan 48824, USA}
\author{B.S.~Acharya} \affiliation{Tata Institute of Fundamental Research, Mumbai, India}
\author{M.~Adams} \affiliation{University of Illinois at Chicago, Chicago, Illinois 60607, USA}
\author{T.~Adams} \affiliation{Florida State University, Tallahassee, Florida 32306, USA}
\author{G.D.~Alexeev} \affiliation{Joint Institute for Nuclear Research, Dubna, Russia}
\author{G.~Alkhazov} \affiliation{Petersburg Nuclear Physics Institute, St. Petersburg, Russia}
\author{A.~Alton$^{a}$} \affiliation{University of Michigan, Ann Arbor, Michigan 48109, USA}
\author{G.~Alverson} \affiliation{Northeastern University, Boston, Massachusetts 02115, USA}
% ...
%
% visitor_addresses.tex                        16 July 2010
%  available symbols are:
%  $\ast, \dag, \ddag, \S, \P, $\|$, $\ast\ast$, \dag\dag, \ddag\ddag ,\#
%
\collaboration{The D0 Collaboration\footnote{with visitors from
%{alton}
$^{a}$Augustana College, Sioux Falls, SD, USA,
%{burdin}
$^{b}$The University of Liverpool, Liverpool, UK,
%{haas,partridge}
$^{c}$SLAC, Menlo Park, CA, USA,
%{juste}
$^{d}$ICREA/IFAE, Barcelona, Spain,
%{luna-garcia}
$^{e}$Centro de Investigacion en Computacion - IPN, Mexico City, Mexico,
%{podesta-lerma}
$^{f}$ECFM, Universidad Autonoma de Sinaloa, Culiac\'an, Mexico,
and 
%{weber}
$^{g}$Universit{\"a}t Bern, Bern, Switzerland.%
%{hooper}
%$^{?}$%Visitor from Bradley University, Peoria, IL, USA.
%{kozminski
%$^{?}$}%Visitor from Lewis University, Romeoville, IL, USA.
%{deceased}
%$^{\ddag}$%Deceased.
}} \noaffiliation
\vskip 0.25cm
  % input Dzero author list
%\setpagewiselinenumbers
%\modulolinenumbers[3]
%\linenumbers

%\widetext \leftline{Version 1.9 for Collaboration Review as of \today} \leftline{Primary
%authors: Xuebing Bu, Pengfei Ding, Liang Han, Yanwen Liu, Hang Yin}
%\leftline{Send comments to d0-run2eb-012@fnal.gov by Monday, June 14, 2010}
\date{July 27, 2010}

\begin{abstract}

  We report the results of a search for $R$ parity violating (RPV) interactions leading
  to the production of supersymmetric sneutrinos decaying into
  $e\mu$ final states using 5.3 fb$^{-1}$ of integrated luminosity
   collected by the D$0$ experiment at the Fermilab
  Tevatron Collider.
  Having observed no evidence for production of $e\mu$ resonances, we set direct bounds
  on the RPV couplings $\lambda^{'}_{311}$ and $\lambda_{312}$
  as a function of sneutrino mass.

\end{abstract}


\maketitle
%%%%%%%%%%%%%%%%%%%%%% intrudction on motivation %%%%%%%%%%%%%%%%%%%%%%%%
In all standard model (SM) interactions, 
baryon number, $B$, and lepton number, $L$, are separately conserved. 
In the supersymmetric (SUSY) extension of the SM, 
$B$ and $L$ violating interactions are generally allowed. 
A new multiplicative quantum number is therefore introduced,
$R=(-1)^{2S+3(B-L)}$, defined in terms of $B$, $L$ and the spin quantum number~$S$,
which distinguishes SM particles ($R=+1$) from their SUSY partners ($R=-1$)
~\cite{RParity}.  If $R$ parity is conserved in the minimal extension of the SM, 
no $B$ and $L$ violating interactions can
occur.  

All SUSY particles are pair-produced if $R$ parity is conserved.
RPV interactions allow single production of SUSY particles
which significantly
reduces the energy required to observe them at a collider.

The most general 
renormalizable gauge invariant $R$ parity violating (RPV)  supersymmetric potential can
be found in~\cite{limits}. The terms in the Lagrangian relevant to this analysis are
\begin{eqnarray}
\lefteqn{
\label{eqn:RPVpotential} {\cal L}_{{\rm RPV}} = 
} \nonumber \\ & & 
-\frac{1}{2} \lambda_{ijk} \left( \tilde{\nu}_{iL}\bar{l}_{kR}l_{jl} 
-  \tilde{\nu}_{jL}\bar{l}_{kR}l_{il} \right)
\nonumber \\
& & 
- \lambda'_{ijk} \left( \tilde{\nu}_{iL}\bar{d}_{kR}d_{jl} \right)+ \mbox{h.c.},
\end{eqnarray}
%\begin{eqnarray}
%\lefteqn{
%\label{eqn:RPVpotential} {W}_{{\rm RPV}} = } \nonumber \\
%& & 
%\frac{1}{2}\varepsilon_{ab} \lambda_{ijk}{L}_{i}^a {L}_{j}^b
% {E}_{k} + \varepsilon_{ab}\lambda^{'}_{ijk} {L}_{i}^a
% {Q}_{j}^b {D}_{k} +  \nonumber \\
%& &   
 %\frac{1}{2}\varepsilon_{\alpha\beta\gamma}\lambda^{''}_{ijk}
 %  {U}_{i}^{\alpha} {D}_{j}^{\beta} {D}_{k}^{\gamma} +
 %\varepsilon_{ab}\delta_{i} {L}_{i}^a {H}_{2}^b,
% \end{eqnarray}
where the indices $i,j,k=1,2,3$ refer to fermion generation; 
$l$ and
$d$ are the SM lepton and down quark fields; $\tilde{\nu}$ is the field of the
SUSY partner of the neutrino, the sneutrino.
These terms lead to the production of a single sneutrino in $d\overline{d}$ scattering.
The search is performed under the hypothesis that only the third generation
sneutrino ($\tilde{\nu}_{\tau}$) is produced and that it is the lightest SUSY particle. All couplings 
apart from $\lambda_{311}^{'}$ and  $\lambda_{312}=\lambda_{321}=-\lambda_{231}=-\lambda_{132}$
are therefore assumed to be zero and the sneutrino decay is determined by the $e\mu$ and
$d\overline{d}$ modes.

In this Letter, we report on a search for resonant production of a
sneutrino decaying into an electron and a muon in $p\overline{p}$
collisions at $\sqrt{s}=1.96$~TeV at the Fermilab Tevatron collider in a data set with an integrated
luminosity of $5.3$~fb$^{-1}$, collected between April 2002 and June 2009.
Observation of this process would provide direct
evidence of physics beyond the SM~\cite{ppem}. 
Previous searches for high-mass $e\mu$ resonances by the 
CDF~\cite{CDFrun2a} and D0~\cite{D0run2a} Collaborations were based on integrated luminosities
of $1.0$ fb$^{-1}$. 
In addition to using a significantly larger data sample, the signal sensitivity
has been improved by increasing the lepton acceptance and by applying
a neural network (NN) selection to distinguish jets
from  electrons.

Indirect two standard deviation bounds on the coupling
constants, under the single coupling dominance assumption with a
degenerate sparticle mass spectrum of $M\equiv
M_{\tilde{\nu}_{\tau}}= 100$~GeV, are given in Ref.~\cite{limits} as
\begin{equation}
\lambda_{311}^{'} \le 0.12,  \hskip 0.cm  \lambda_{312} \le 0.07.
\label{eqn:limits}
\end{equation}
The sneutrino production cross section is determined by these two couplings and
the sneutrino mass $M_{\tilde{\nu}_{\tau}}$. 
The final state is characterized by an electron and a muon, both
of which are well-isolated and have high transverse momentum, $p_T$,
of approximately $M_{\tilde{\nu}_{\tau}}/2$, and by a peak in the invariant
$e\mu$ mass at $M_{\tilde{\nu}_{\tau}}$. The dominant SM
background processes for this event topology are the production of $Z/\gamma^*\rightarrow\tau\tau$,
dibosons ($WW/WZ/ZZ$),  $t\bar{t}$ pairs, and
$W$ bosons in association with jets, where a jet is misidentified as a lepton.

The D0 detector~\cite{d0_detector}
comprises a central tracking system in a $2$~T superconducting
solenoidal magnet, surrounded by a central preshower detector (CPS), a
calorimeter, and a muon system. The tracking system, a
silicon microstrip tracker (SMT) and a scintillating fiber tracker
(CFT), provides coverage for charged particles in the pseudorapidity
range $|\eta|<3$~\cite{d0_eta}. The CPS is located immediately
before the inner layer of the calorimeter and is formed of
approximately one radiation length of lead absorber followed by
three layers of scintillating strips. The calorimeter consists of a
central calorimeter (CC) covering  up to $|\eta|\approx 1.1$, and two end
caps (EC) extending coverage to $|\eta|\approx 4.2$. Each 
consists of an inner electromagnetic (EM) section, followed by a
hadronic section. The EM calorimeter has four longitudinal
layers and transverse segmentation of $0.1\times 0.1$ in $\eta-\phi$
space, except in the third layer, where it is $0.05\times0.05$. 
The muon system resides beyond the calorimeter and consists of a layer
of tracking detectors and scintillation trigger counters before $1.8$~T
iron toroidal magnets, followed by two similar layers after the
toroids. The coverage of the muon system is $|\eta|<2$.
The data acquisition system consists of a three-level
trigger, designed to accommodate the high instantaneous luminosity.
For final states containing an electron with $p_T>30$~GeV, the
trigger efficiency is close to $100\%$. 

To simulate signal kinematics in the D0 detector, parton level
signal events are generated using the {\sc {comphep}}~\cite{CompHEP} 
leading order Monte Carlo generator and then processed through {\sc
{pythia}}~\cite{PYTHIA} to include parton showering, hadronization,
and particle decays. 
SM background processes are generated with {\sc pythia},
except for the $W$+jets inclusive samples, which
are generated with {\sc alpgen} \cite{alpgen} and {\sc
{pythia}} for parton showering. 
All signal and background simulations use the CTEQ6L1~\cite{CTEQ} 
parametrization of the parton distribution functions.

We use next-to-next-to-leading order (NNLO) cross sections for
Drell-Yan $Z/\gamma^*\to \ell \ell$ $(\ell=e,\tau,\mu)$ processes~\cite{Drell-Yan-X} and
NLO cross sections for diboson~\cite{diboson-X},
$t\bar{t}$~\cite{ttbar-X}, and $W+$jets production~\cite{wjet-x}.
All signal and background events are processed with a detailed {\sc
{geant}}-based D$0$ detector simulation~\cite{D0-simulation} and
are corrected for trigger effects and for the differences in the
reconstruction efficiencies in the simulation compared to those in data.

Electrons are selected by requiring an EM cluster
in the CC or in either EC  with transverse energy $E_T > 30$~GeV
 within a cone of radius ${\cal R} =\sqrt{(\Delta \phi)^2 + (\Delta \eta)^2} = 0.2$.
 The EM cluster in the CC must be in the range $|\eta|<1.1$ and in the EC
 in the range $1.5<|\eta|<3.2$.
 At least $97\%$ of the cluster energy must be
 deposited in the EM section of the calorimeter and
 the energy must be isolated in the calorimeter, $[{E_{\text{tot}}(0.4) -
E_{\text{EM}}(0.2)}]/{E_{\text{EM}}(0.2)}<0.07$, where
$E_{\text{tot}}({\cal R})$ and $E_{\text{EM}}({\cal R})$ are the total energy and the energy in the EM
section, respectively, within a cone of radius ${\cal R}$ around the electron direction.
A track must point to the EM cluster for all electron candidates. 
A multi-variable likelihood discriminant, which includes information from the
 spatial track match, must be consistent with that for an electron in the CC.
 An NN is trained using information
 from the tracker, the calorimeter, and the CPS to further reject background from
 jets misidentified as electrons. The electron
 must also be spatially separated from reconstructed muons.

 For the muon candidate, we require that the associated central track $p_T$
 exceeds $25$~GeV and that the time measured for hits in the muon scintillation counters is consistent
with an interaction originating from a $\ppbar$ collision. The
central track fit must have $\chi^2/\mbox{ndf}<4$ and  the distance
of closest approach (dca) of the track to the beam spot in the plane transverse to the beam direction should be
less than $0.02$~cm if the track has SMT hits and less than $0.2$~cm otherwise.
 The sum of the  transverse energy of calorimeter cells in the annulus  $0.1 <  {\cal R} < 0.4$ around
 the muon direction must  be
 less than $2.5$~GeV, and the sum of the  transverse momentum of all tracks besides the muon track within
 ${\cal R} = 0.5$ must be less than $2.5$~GeV.
 
  \begin{figure}[htbp]
  \includegraphics[scale=0.43]{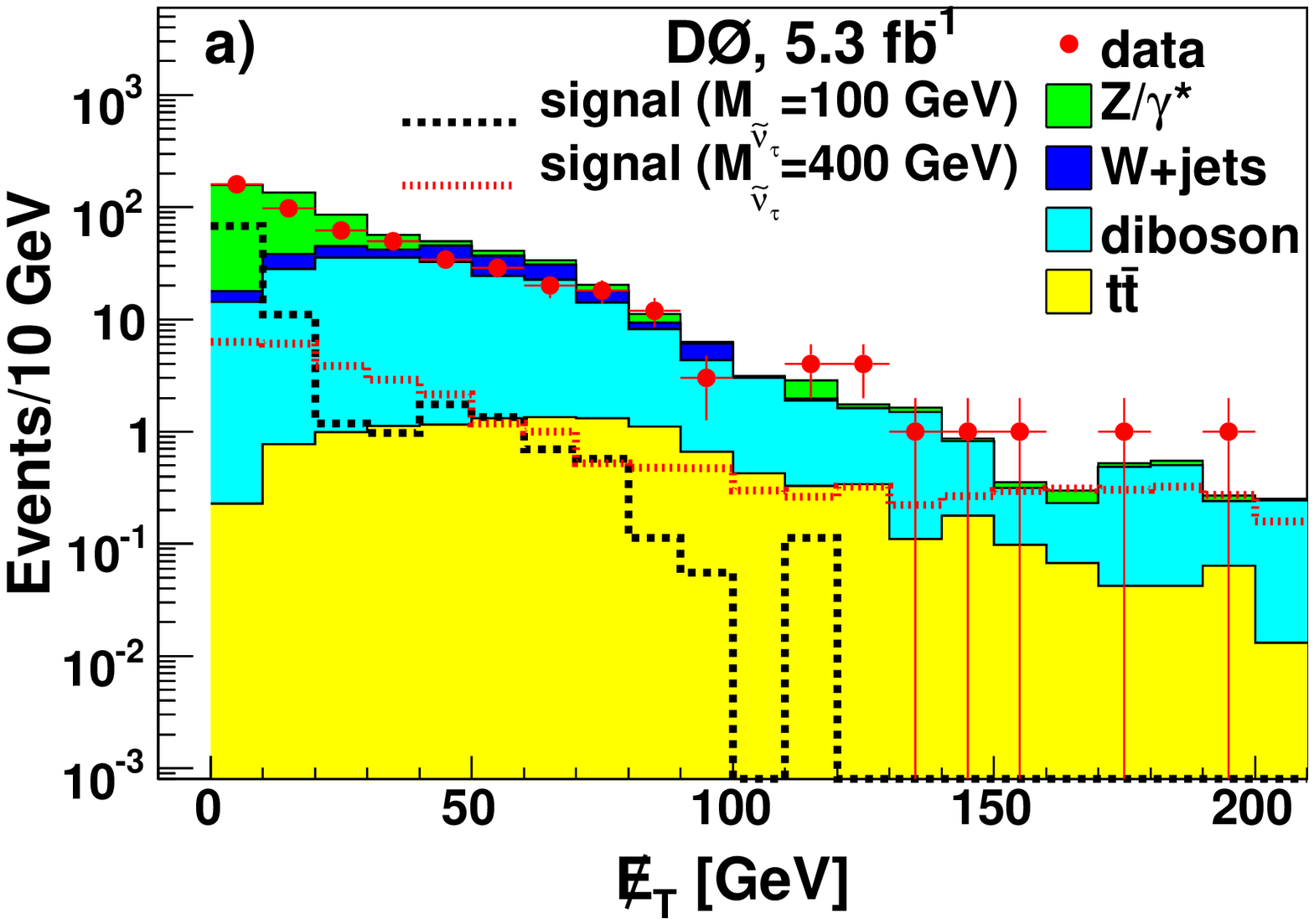}
  \includegraphics[scale=0.43]{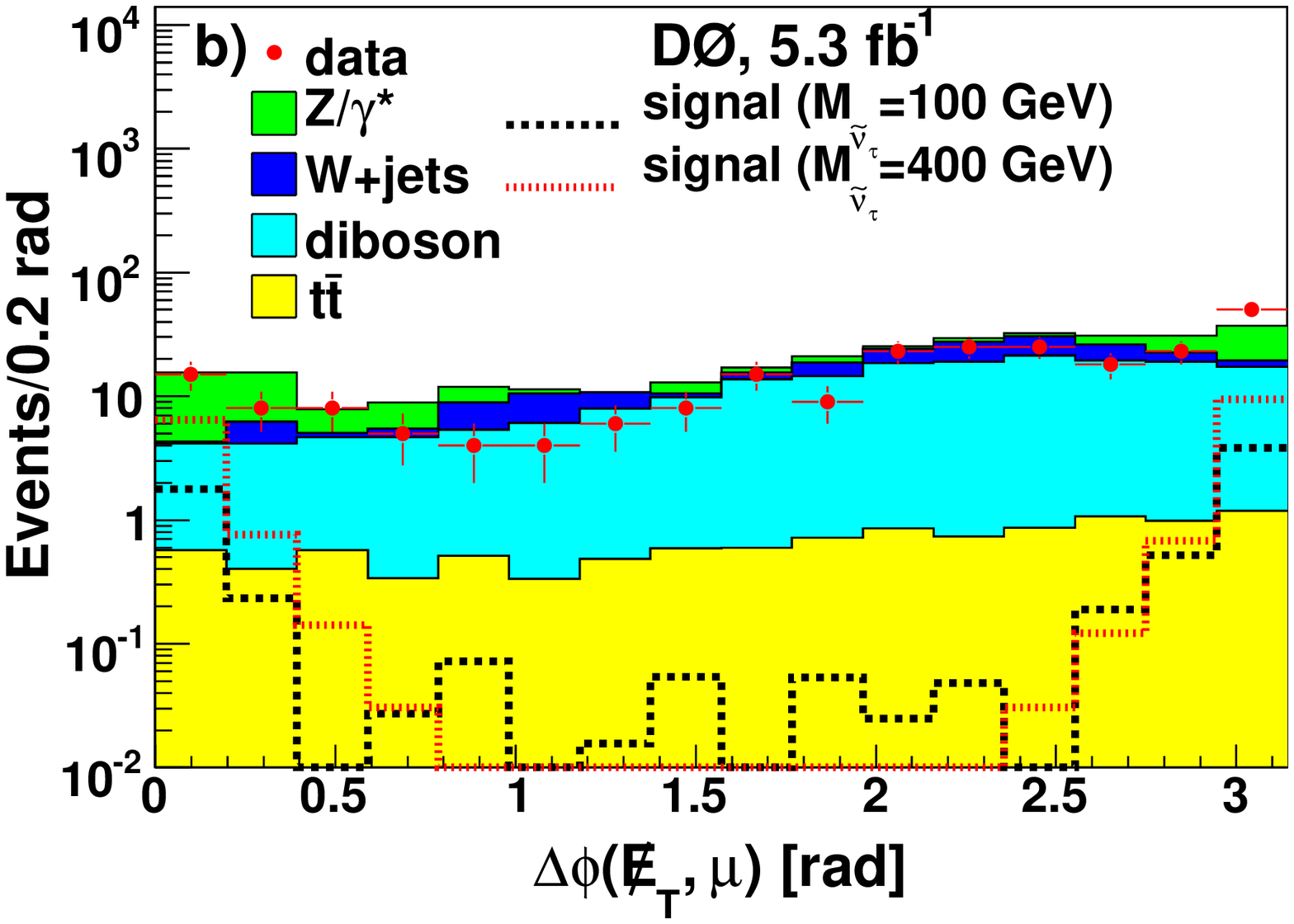}
  \caption{\label{fig:MET02noveto}  (color online).
Distributions of (a) $\MET$ and (b) $\Delta\phi(\MET,\mu)$ between
the direction of the $\MET$ and the muon for $\slashed{E}_T>20$~GeV.
The signal simulation is shown for $M_{\tilde{\nu}_{\tau}}=100$~GeV  ($400$~GeV) 
 and $\sigma\times\mbox{BR}=40$~fb ($12.5$~fb).  }
\end{figure}
\begin{figure}[htbp]
  \includegraphics[scale=0.43]{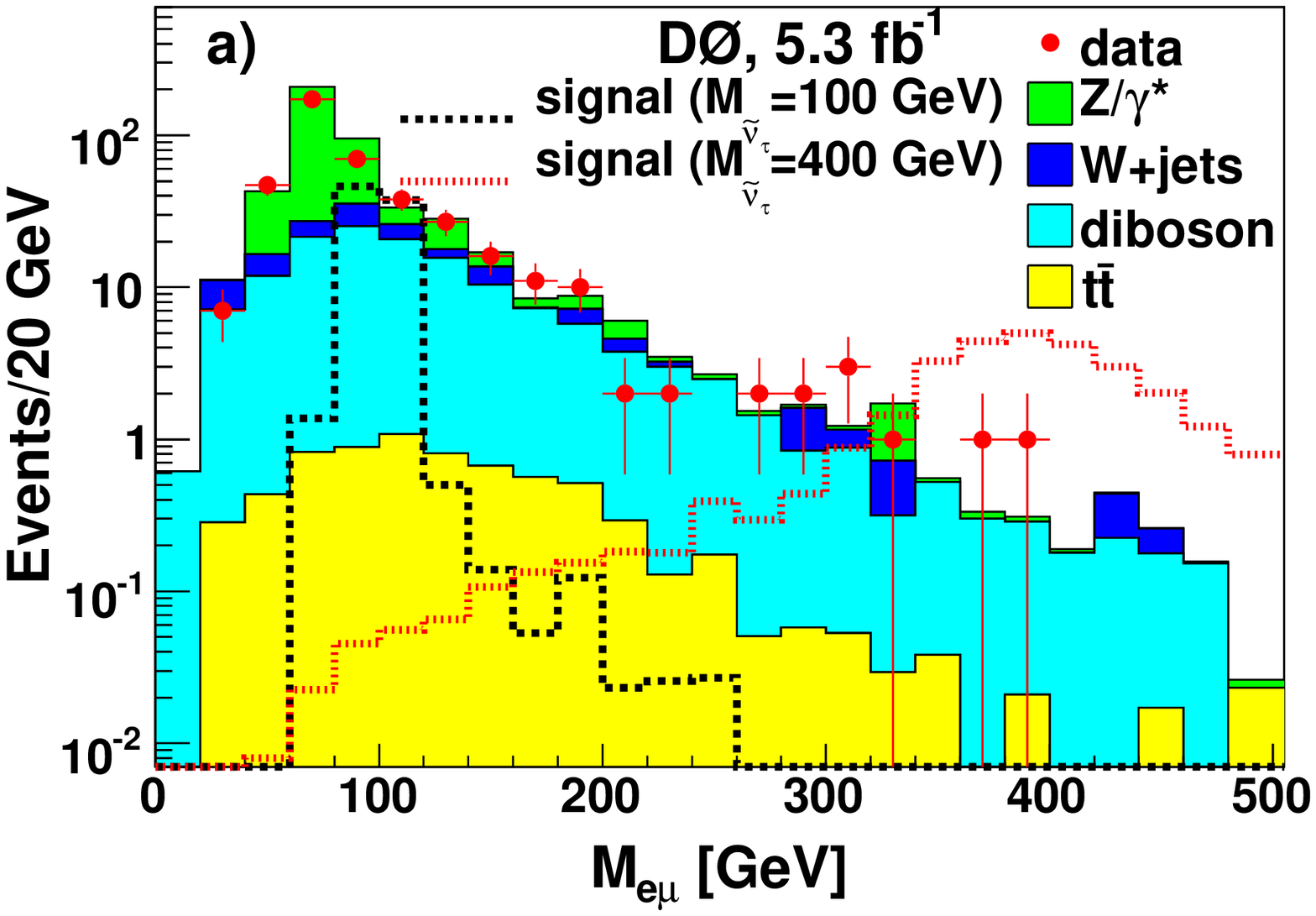}
  \includegraphics[scale=0.43]{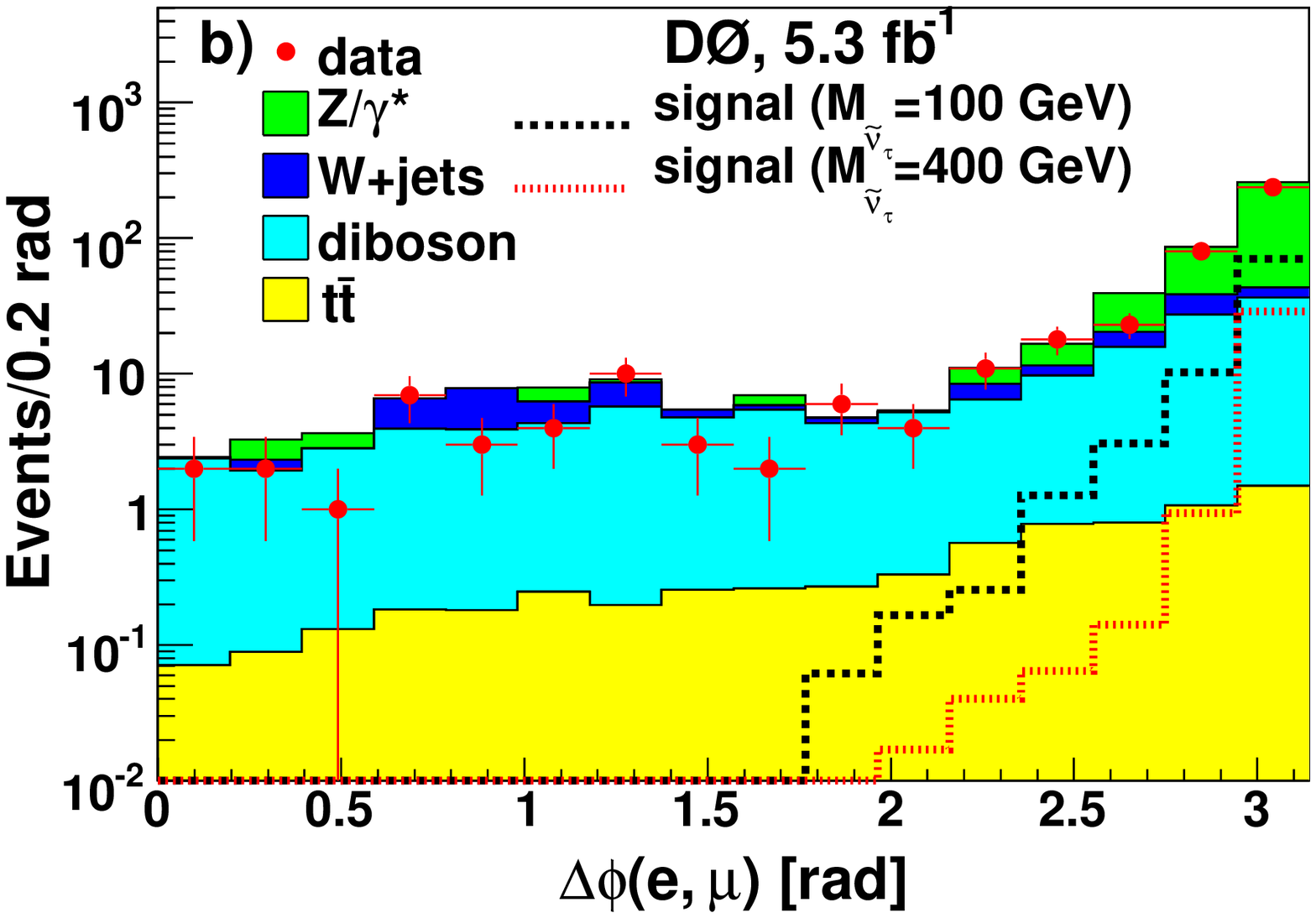}
  \caption{\label{fig:dphiem} (color online). Distributions of 
  (a)  $M_{e\mu}$ and (b) $\Delta\phi(e,\mu)$ for data
 compared to the sum of all background processes after the final event selection,
rejecting, in addition, events with
$\slashed{E}_T>20$~GeV and $0.7<\Delta\phi(\slashed{E}_T,\mu)<2.3$.
The signal simulation is shown for $M_{\tilde{\nu}_{\tau}}=100$~GeV  ($400$~GeV) 
 and $\sigma\times\mbox{BR}=40$~fb ($12.5$~fb). }
\end{figure}

Events are required to have exactly one high $p_T$ isolated
electron candidate and one high $p_T$ isolated muon candidate. There is no
requirement on the charge of the two leptons.
Furthermore, events are only considered if the primary vertex is reconstructed within $60$~cm of the
center of the D$0$ detector in the $z$ coordinate along the beam axis and if
the difference between the $z$ coordinates of the muon and the electron at the dca is
less than $2$~cm.
Background from $Z/\gamma^*\rightarrow\tau\tau$
decays is heavily suppressed by the high $p_T$ requirement for the two leptons. 
To reduce the background from $t\bar{t}$ production, 
events are rejected if they have at least one jet with $p_T > 25$~GeV in the range
$|\eta| < 2.5$, where jets are reconstructed using an
iterative seed-based cone algorithm~\cite{jet_definition}. 
Figure~\ref{fig:MET02noveto}(a) shows the missing transverse energy ($\MET$) distribution and Fig.~\ref{fig:MET02noveto}(b)
 the distribution of  azimuthal angles between the $\MET$ and the muon direction, $\Delta\phi(\slashed{E}_T,\mu)$,
 for events with $\MET>20$~GeV. Good agreement between data and the total SM predictions is observed.
Signal events have low $\MET$, but
due to the limited muon momentum resolution, some $\MET$ is expected in $e\mu$ signal
events that is either pointing in the muon direction or opposite to it. This is observed
in Fig.~\ref{fig:MET02noveto}(b) for signal events and for the topologically similar Drell-Yan
process. Events with
$\slashed{E}_T>20$~GeV are therefore rejected only if $0.7<\Delta\phi(\slashed{E}_T,\mu)<2.3$.

The resulting distribution of the electron and muon invariant
mass $M_{e\mu}$ and the azimuthal angle  $\Delta\phi(e,\mu)$ between the electron and muon direction are
shown in Fig.~\ref{fig:dphiem}.
The kinematic variables of
the $e\mu$ final state are well described by the sum of the SM background
contributions.
The background contributions and the number of selected candidates
are summarized in Table~\ref{tab:sum}. About $80\%$ of the Drell-Yan background is due
to $Z/\gamma^*\to\tau\tau$ events.

\begin{table}
  \caption{\label{tab:sum} The numbers of selected events in data and
    estimated background contributions for an integrated luminosity of $5.3$~fb$^{-1}$ with
    their total uncertainties.}
  \begin{tabular}{cc}
    \hline \hline
  Process & Number of events\\
    \hline
    Drell-Yan ($Z/\gamma^*$)          & $254 \pm 26$\\
    Diboson  ($WW,WZ,ZZ$)           & $116 \pm 12$ \\
    $t\bar{t}$                                       & $\phantom{0} 5.8 \pm 1.0$ \\
    $W$+jets                                      & $34.1 \pm 5.9$ \\
    \hline
    Total background                        & $410 \pm 38$\\
    \hline
    Data                                               & 414\\
    \hline
   \hline
  \end{tabular}
\end{table}

There are 414 candidate events found in the data. The expectation from 
SM processes is $410 \pm 38$ events, where the uncertainty includes the
statistical uncertainty of the MC, the systematic uncertainties from the integrated
luminosity ($6.1\%$), reconstruction and trigger efficiencies (0.5\%), which are all taken to be fully correlated
between the background sources, and the uncertainties
on the cross sections ($Z/\gamma^*\rightarrow {\ell}{\ell}$ 3.5\%, $t\bar{t}$ 14.8\%,  diboson
production $2.7\%-6.6\%$, and $W+$jets $8.5\%$). 
Additional PDF uncertainties on the signal acceptance are estimated from the 
CTEQ6.1M eigenvector PDF sets and lie in the range $0.4\% - 0.6\%$, depending on
$M_{\tilde{\nu}_{\tau}}$. 

\begin{figure}
  \includegraphics[scale=0.43]{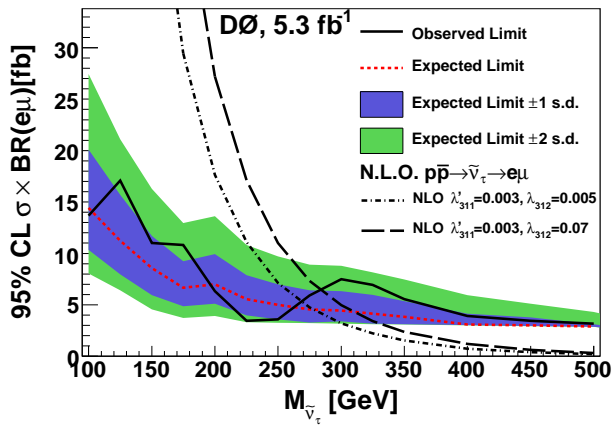}
  \caption{\label{fig:xsection} (color online). The observed and median expected upper limits on
$\sigma\times\mbox{BR}$ for the process $\ppbar\rightarrow\tilde{\nu}_{\tau}+X\rightarrow e\mu+X$ as 
function of  $M_{\tilde{\nu}_{\tau}}$. The median expected limits are shown together
with the $\pm 1$ and $\pm 2$ standard deviation bands. The theoretical cross sections
for $\lambda'_{311}=0.003$ and $\lambda_{312}=0.005$ and $0.07$ are also shown.}
\end{figure}

The $M_{e\mu}$ distribution is used to calculate an upper limit on the production
cross section multiplied by the branching ratio,
$\sigma\times\mbox{BR}$, for the process
$\ppbar\rightarrow\tilde{\nu}_{\tau}+X\rightarrow e\mu+X$ with a
modified frequentist (CL$_s$) method~\cite{lhood-ratio}, under the
assumption that the total width of the produced resonance is much narrower than the detector
resolution. 
The observed cross section upper limits as a function of the $M_{\tilde{\nu}_{\tau}}$
hypothesis are shown in Fig.~\ref{fig:xsection}, together with
the median expected limits.

\begin{figure}[ht]
  \includegraphics[scale=0.43]{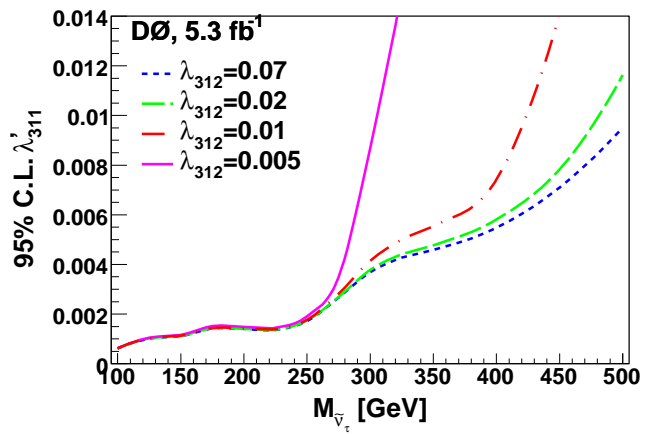}
  \caption{\label{fig:couplings}  (color online). The 95\% C.L.~observed upper limits on
        $\lambda^{\prime}_{311}$ for four values of $\lambda_{312}$ as a function of  $M_{\tilde{\nu}_{\tau}}$.}
\end{figure}

These limits are translated into upper limits on couplings as a function of $M_{\tilde{\nu}_{\tau}}$
using the theoretical signal cross section~\cite{ppem}.
A mass dependent $K$-factor, ranging from $1.64$ at $M_{\tilde{\nu}_{\tau}}=100$~GeV
to $1.29$ at $M_{\tilde{\nu}_{\tau}}=500$~GeV, is applied to the cross section
to include next-to-leading order (NLO) QCD corrections~\cite{ppemQCD}.
The limits are obtained by fixing one of the coupling constants and then setting the upper limit on the other for
different $M_{\tilde{\nu}_{\tau}}$. In Fig.~\ref{fig:couplings},
the observed upper limits on $\lambda^{\prime}_{311}$ for four
assumed values of $\lambda_{312}$ are shown.
For $M_{\tilde{\nu}_{\tau}}=100$~GeV and $\lambda_{312}\le 0.07$, 
couplings $\lambda^{\prime}_{311}>6.2\times10^{-4}$ are excluded at the $95\%$~C.L. 

In summary, we have searched for a high mass, narrow $e\mu$ resonance in
D$0$ data corresponding to an integrated luminosity of $5.3$~fb$^{-1}$.
A total of $414$ $e\mu$ events is selected in data, in agreement with the predicted
number of SM events, $410 \pm 38$. The kinematic distributions 
are well described by the SM predictions. 
The upper limits on the production cross section for such a resonance
are about a factor of 5 lower than for our previous
result~\cite{D0run2a} and about a factor 6 lower for $M_{\tilde{\nu}_{\tau}}=500$~GeV
compared to a search performed in the $e\mu$ channel by the CDF Collaboration with
$1$~fb$^{-1}$ of integrated luminosity~\cite{CDFrun2a}.
We have also derived limits on the parameters of a SUSY model that
predicts a sneutrino resonance  decaying into $e\mu$ via 
$R$ parity violating production and decay of sneutrinos.

% acknowledgement.tex                             16 July 2010
%
We thank the staffs at Fermilab and collaborating institutions,
and acknowledge support from the
DOE and NSF (USA);
CEA and CNRS/IN2P3 (France);
FASI, Rosatom and RFBR (Russia);
CNPq, FAPERJ, FAPESP and FUNDUNESP (Brazil);
DAE and DST (India);
Colciencias (Colombia);
CONACyT (Mexico);
KRF and KOSEF (Korea);
CONICET and UBACyT (Argentina);
FOM (The Netherlands);
STFC and the Royal Society (United Kingdom);
MSMT and GACR (Czech Republic);
CRC Program and NSERC (Canada);
BMBF and DFG (Germany);
SFI (Ireland);
The Swedish Research Council (Sweden);
and
CAS and CNSF (China).


\begin{thebibliography}{99}
\bibitem{d0_eta}
  The polar angle $\theta$ and the azimuthal angle $\phi$ are defined
  with respect to the positive z axis, which is along the proton beam
  direction. Pseudorapidity is defined as $\eta = -\ln[\tan(\theta/2)]$.
\end{thebibliography}
\end{document}